\documentstyle[12pt,twoside,fleqn,espcrc1,epsf]{article}

\newcommand{\AmS}{{\protect\the\textfont2
  A\kern-.1667em\lower.5ex\hbox{M}\kern-.125emS}}

\title{Mapping the QCD Phase Diagram}

\author{Krishna Rajagopal\address{Center for Theoretical Physics,
MIT, Cambridge, MA 02139, USA}\thanks{Those features 
of the QCD phase diagram that I sketch here 
which I have helped to understand were understood over the
course of extended and very enjoyable collaborations with
M. Alford, J. Berges, E. Shuryak, M. Stephanov and F. Wilczek,
to all of whom I am grateful. I am also grateful to R. Jaffe, 
G. Roland, T. Sch\"afer and D. Son for helpful discussions,
to B. Berdnikov and E. Shuster for reading the manuscript
and to
the organizers of Quark Matter '99 for the opportunity to
visit Torino and the invitation to give this talk.
Research supported in part by
a DOE OJI Award, by the A. P. Sloan Foundation and by the DOE
under agreement DE-FC02-94ER40818. Preprint MIT-CTP-2895.}}

\begin{document}
\maketitle



The QCD vacuum in which we live, which has the familiar hadrons
as its excitations, is but one phase of QCD, and far from
the simplest one at that. One way to better understand
this phase and the nonperturbative dynamics of QCD more generally
is to study other phases and the transitions between phases.
We are engaged in a voyage
of exploration, mapping the QCD phase diagram as a function
of temperature $T$ and baryon number chemical potential $\mu$.
Because of asymptotic freedom, the high temperature and
high baryon density phases of QCD are more simply and more
appropriately described in terms of quarks and gluons as
degrees of freedom, rather than hadrons. The chiral symmetry
breaking condensate which characterizes the vacuum phase
melts away.  At high densities, quarks form Cooper pairs and new
condensates develop.  The formation
of such superconducting
phases \cite{Barrois,BailinLove,ARW1,RappETC} 
requires only weak attractive interactions; these phases may nevertheless
break chiral symmetry \cite{CFL} 
and have excitations which are indistinguishable
from those in a confined phase \cite{CFL,SW1,ABR,SW2}.   These 
cold dense quark matter phases may arise in the centers of
neutron stars; mapping this region of the phase diagram
will require an interplay between theory and neutron star
phenomenology. The goal of the experimental heavy ion
physics program is to explore and map the higher temperature
regions of the diagram. Recent theoretical developments
suggest that a key qualitative feature, namely a critical
point which in a sense defines the landscape
to be mapped, may be within reach of discovery and analysis
by the CERN SPS, if data is taken at several different
energies \cite{SRS1,SRS2}. The discovery of the critical point
would in a stroke transform the map of the QCD phase
diagram which we sketch below from one based only on
reasonable inference from universality, lattice gauge theory
and models into one with a solid experimental basis.

\section{The Critical Point on the Map}

We begin our walk through the phase diagram at zero
baryon number density, with a brief review \cite{rajreview} 
of the phase changes
which occur as a function of temperature.
That is, we begin by restricting ourselves to 
the vertical axis in Figures 1 through 4.  This slice of
the phase diagram was explored by the early universe
during the first tens of microseconds after the big bang
and can be studied in lattice simulations.  As heavy ion collisions
are performed at higher and higher energies, they create plasmas
with a lower and lower baryon number to entropy ratio and therefore
explore regions of the phase diagram 
closer and closer to the vertical axis.  

\begin{figure}[t]
\begin{center}
\vspace{-0.1in}
\hspace*{0in}
\epsfysize=2.5in
\epsffile{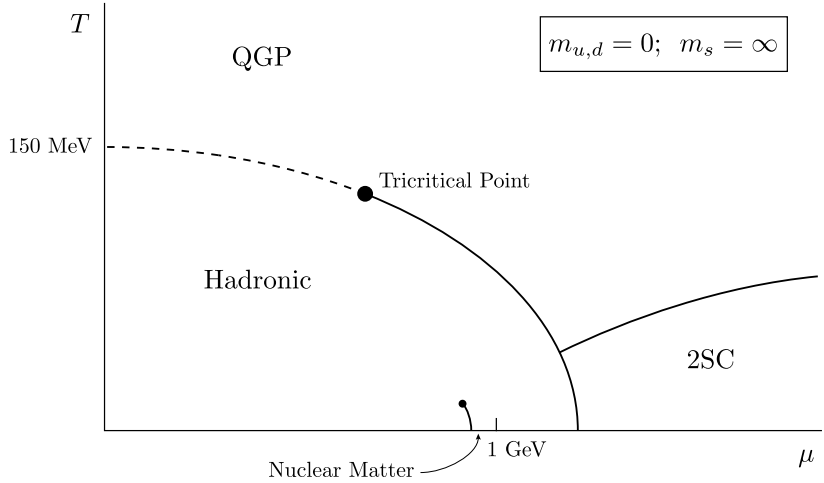}
\end{center} \label{fig1}
\vspace*{-0.5in}
\caption{QCD Phase diagram for two massless quarks. Chiral symmetry
is broken in the hadronic phase and is restored elsewhere in the 
diagram. The chiral phase transition changes from second to
first order at a tricritical point. At high densities and low
temperatures, we find a superconducting phase in which up and down
quarks with two out of three colors pair and form a condensate.
The transition between this 2SC phase and the QGP phase 
is likely first order. The transition on the horizontal
axis between the hadronic and 2SC phases is first 
order. The transition between a nuclear matter ``liquid''
and a gas of individual nucleons is also marked; it ends
at a critical point at a 
temperature of order 10 MeV, characteristic of the forces
which bind nucleons into nuclei.}
\vspace{-0.2in}
\end{figure}
In QCD with
two massless quarks ($m_{u,d}=0$; $m_s=\infty$; Figure 1)
the phase transition at which chiral symmetry is restored 
is likely second order and belongs to the universality
class of $O(4)$ spin models in three dimensions \cite{piswil}.
Below $T_c$, chiral symmetry is broken and there are three
massless pions.  At $T=T_c$, there are four massless degrees
of freedom: the pions and the sigma. Above $T=T_c$, the pion
and sigma correlation lengths are degenerate and finite.
In nature, the light quarks are not massless.  Because
of this explicit chiral symmetry breaking,
the second order phase transition is replaced by an 
analytical crossover: physics changes dramatically but smoothly in the 
crossover region, and no correlation length diverges.
Thus, in Figure 2, there is no sharp boundary on the 
vertical axis separating the low temperature hadronic world
from the high temperature quark-gluon plasma.  
This picture is consistent with present lattice 
simulations \cite{latticereview},
which suggest $T_c\sim 140-170$ MeV \cite{latticeTc}.
\begin{figure}[t]
\begin{center}
\vspace{-0.1in}
\epsfysize=2.5in
\hspace*{0in}
\epsffile{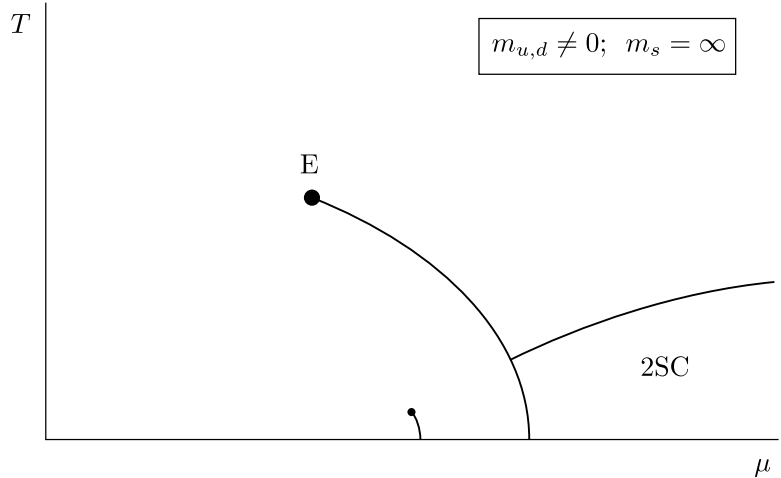}
\end{center} \label{fig2}
\vspace*{-0.5in}
\caption{QCD phase diagram for two light quarks. Qualitatively as in Figure 1,
except that the introduction of light quark masses  turns the
second order phase transition into a smooth crossover. 
The tricritical point becomes the critical endpoint $E$, which
can be found in heavy ion collision experiments.}
\vspace{-0.1in}
\end{figure}

Arguments based on a variety of 
models \cite{NJL,steph,ARW1,RappETC,bergesraj,stephetal}
indicate that the transition as a function of $T$ is first order at 
large $\mu$.  
This suggests that the
phase diagram features a critical point $E$ at which
the line of first order phase transitions present for 
$\mu>\mu_E$ ends, as shown in Figure 2.\footnote{If
the up and down quarks were massless, $E$ would
be a tricritical point \cite{lawrie}, at which the first
order transition becomes second order. See Figure 1.}
At $\mu_E$, the phase transition is second order
and is in the Ising universality class \cite{bergesraj,stephetal}.
Although the
pions remain massive, the correlation length in the $\sigma$ channel
diverges due to universal long wavelength fluctuations
of the order parameter.
This results in characteristic signatures,
analogues of critical opalescence in the sense that they
are unique to collisions which freeze out near the
critical point, which
can be used to discover $E$ \cite{SRS1,SRS2}.   

Returning to the $\mu=0$ axis,
universal arguments \cite{piswil}, again backed by lattice 
simulation \cite{latticereview},
tell us that if the strange quark were as light as the
up and down quarks, the transition would be first order,
rather than a smooth crossover.  
This means that if one could dial
the strange quark mass $m_s$, one would find a critical
$m_s^c$ at which the transition as a function of temperature
is second order \cite{rajwil,rajreview}. 
Figures 2, 3 and 4 are drawn
for a sequence of decreasing strange quark masses. Somewhere
between Figures 3 and 4, $m_s$ is decreased below $m_s^c$ and
the transition on the vertical axis becomes first order.
The value of $m_s^c$ is an open question,
but lattice simulations suggest that it is about half the
physical strange quark mass \cite{columbia,kanaya}. 
These results are not yet conclusive \cite{oldkanaya}  but
if they are correct then the phase
diagram in nature is as shown in Figure 3,  and the phase transition
at low $\mu$ 
is a smooth crossover.
\begin{figure}[t]
\begin{center}
\vspace{-0.1in}
\epsfysize=2.5in
\hspace*{0in}
\epsffile{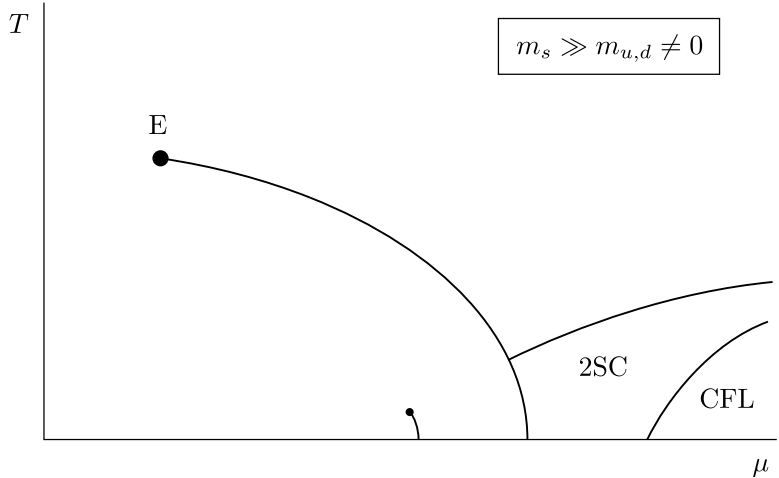}
\end{center} \label{fig3}
\vspace*{-0.5in}
\caption{QCD phase diagram for two light quarks and a strange quark
with a mass comparable to that in nature.
The presence of the strange quark shifts
$E$ to the left,  as can be seen by comparing with Figure 2. 
At sufficiently high density, cold quark matter is necessarily 
in the CFL phase in which quarks of all three colors and
all three flavors form Cooper pairs. The diquark condensate in
the CFL phase breaks chiral symmetry, and this
phase has the same symmetries as baryonic matter which is
dense enough that the nucleon and hyperon densities are 
comparable. The phase transition
between the CFL and 2SC phases is first order.}
\vspace{-0.15in}
\end{figure}
\begin{figure}[t]
\begin{center}
\vspace{-0.1in}
\epsfysize=2.5in
\hspace*{0in}
\epsffile{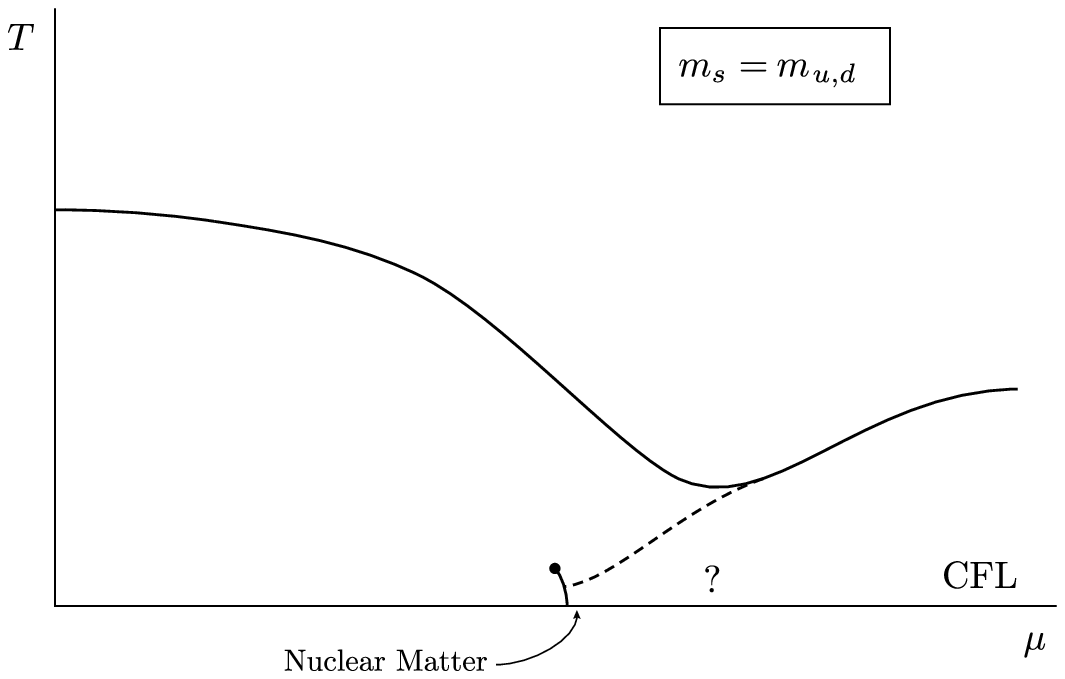}
\end{center} \label{fig4}
\vspace*{-0.5in}
\caption{QCD phase diagram for three quarks which
are degenerate in mass and which are either massless or light.
The CFL phase
and the baryonic phase have the same symmetries and 
may be continuously connected. The dashed line 
denotes the critical temperature at which baryon-baryon (or
quark-quark) pairing vanishes; this dashed line may actually
represent more than one phase transition.
Chiral symmetry is 
broken everywhere below the solid line, which is a first order
phase transition. The question mark serves to remind
us that although no transition is required in this region,
transition(s) may nevertheless arise.  For quark
masses as in nature, the high density region of the map
may be as shown in Figure 3 or may be closer to that shown here, albeit
with additional transitions associated with the onset
of nonzero hyperon density and the breaking of $U(1)_S$ \cite{ABR}.}
\vspace{-0.15in}
\end{figure}

These observations fit together in a simple
and elegant fashion.
If we
could vary  $m_s$, what we would find is that as $m_s$
is reduced from infinity to $m_s^c$, the critical
point $E$ in the $(T,\mu)$ plane moves toward the $\mu=0$
axis \cite{SRS1}.  This is shown in Figures 2-4.
In nature, $E$ is at some nonzero $T_E$ and $\mu_E$.
When $m_s$ is reduced to $m_s^c$, between Figure 3 and Figure 4,
$\mu_E$ reaches zero.
Of course, experimentalists cannot vary $m_s$.  They
can, however, vary $\mu$.  The AGS, with beam energy 11 AGeV
corresponding to
$\sqrt{s}=5$ GeV, creates fireballs which freeze out
near $\mu\sim 500-600$ MeV \cite{PBM}.  
When the SPS runs with $\sqrt{s}=17$ GeV
(beam energy 158 AGeV),
it creates fireballs which freeze out near $\mu\sim 200$ MeV \cite{PBM}.
By dialing $\sqrt{s}$ and thus $\mu$, experimenters can find
the critical point $E$.  

\section{Discovering the Critical Point}

The map of the QCD phase diagram which we have sketched
so far is simple, coherent and consistent with all we know
theoretically; the discovery of the critical point would
provide an experimental foundation for the central
qualitative feature of the landscape.
This discovery would in addition
confirm that in higher energy heavy ion collisions and in
the big bang, the QCD phase transition is a smooth crossover.
Furthermore, the 
discovery of collisions
which create matter that
freezes out near $E$ would imply that 
conditions above the transition existed prior to freezeout,
and would thus make it much easier to interpret the results of other
experiments which seek signatures which probe the 
partonic matter created early in the collision. 

We theorists must clearly do as much as we can
to tell experimentalists {\it where} and {\it how} to find $E$.
The ``where'' question, namely the question of predicting
the value of $\mu_E$ and thus suggesting the $\sqrt{s}$ to
use to find $E$, is much harder for us to answer.   One of the things
which is intrinsic to the picture we have described is
that $\mu_E$ is sensitive to the mass of the strange quark,
and therefore particularly hard to predict.
Crude models suggest that $\mu_E$ could be $\sim 600-800$ MeV
in the absence of the strange quark \cite{bergesraj,stephetal}; 
this in turn suggests that
in nature $\mu_E$ may have of order half this value, and may therefore
be accessible at the SPS if the SPS 
runs with $\sqrt{s}<17$ GeV.   However, at present theorists cannot
predict the value of $\mu_E$ even to within a factor of two.
The SPS can search a significant fraction of the parameter
space; if it does not find $E$, it will then be up to 
the RHIC experiments to map the $\mu_E< 200 $ MeV region.

The other half of the 
``where'' question is the question of how close does
one have to come to $E$ to detect its presence.  That is,
how big steps in $\sqrt{s}$ can one safely take and still
be reasonably confident of discovering $E$ if it is there
to be found?  To answer this,
we must estimate $\Delta \mu$, the width in $\mu$ 
of the region centered at $\mu_E$ 
within which the correlation length is longer than $2$ fm,
and thus detectable effects of $E$ arise \cite{SRS2}.
Here again, only crude models are available.  Analysis
within the toy model of Ref. \cite{bergesraj} suggests that 
in the absence of the strange quark, $\Delta \mu \sim 120$ MeV
for $\mu_E\sim 800$ MeV.  Similar results can be obtained \cite{misha}
within a random matrix model \cite{stephetal}.  It is likely over-optimistic
to estimate $\Delta \mu \sim 120$ MeV when the effects
of the strange quark are included and $\mu_E$ itself is significantly
reduced.  A conservative estimate would be to use the models
to estimate that $\Delta \mu/\mu_E \sim 15\%$ in an infinite system.
Finite size effects must increase $\Delta \mu/\mu_E$. As
a reasonable estimate, the best available at present
for use in planning experimental strategy, one may take
$\Delta \mu/\mu_E \sim 20-30\%$.
This suggests that if experiments could be 
done at about four energies between AGS energy
($\sqrt{s}=5$) and maximum SPS energy ($\sqrt{s}=17$ GeV), 
there would be a good chance of finding $E$ if it lies within the range
$200 {\rm ~MeV} < \mu_E < 600 {\rm ~MeV}$.   An SPS run 
at $\sqrt{s}=9$ GeV (beam energy 40 AGeV)
is already planned. This data,
together with that already taken at  $\sqrt{s}=17$ GeV,
together with data from two (or perhaps only one) additional beam energies 
in between would allow the SPS to search for the critical point
over a substantial range of parameter space.  There is therefore
a strong scientific argument for an 80 AGeV run.  If runs at 
two energies between 40 AGeV and 158 AGeV were
possible, that would be ideal.

It should be clear by now that although we are trying
to be helpful with the ``where'' question, we are not very
good at answering it quantitatively.  This question can only
be answered convincingly by an experimental discovery.
What we as theorists {\it can} do reasonably
well is to answer the ``how'' question, thus
enabling experimenters to answer ``where''.  
This is the goal of a recent paper by Stephanov, myself
and Shuryak \cite{SRS2,SRSshortversions}.
The signatures we have proposed are based
on the fact that $E$ is a genuine thermodynamic singularity
at which susceptibilities diverge and the order parameter
fluctuates on long wavelengths. The resulting signatures
are {\it nonmonotonic} as a function of $\sqrt{s}$: as
this control parameter is varied, we should see the signatures
strengthen and then weaken again as the critical point is
approached and then passed.   

The simplest observables we
analyze are the event-by-event fluctuations of the 
mean transverse momentum of the charged particles
in an event, $p_T$, and of the total charged multiplicity in an event, $N$.
We calculate the magnitude of the effects of critical
fluctuations on these and other observables, making
predictions which, we hope, will allow experiments to find $E$.
As a necessary prelude, we analyze the contribution of
noncritical thermodynamic fluctuations. 
We compare the
noncritical fluctuations of an equilibriated resonance gas
to the fluctuations measured by NA49 at $\sqrt{s}=17$ GeV \cite{NA49}. 
The observed
fluctuations are as perfect Gaussians as the data statistics
allow, as expected for freeze-out from a system in thermal equilibrium.
The width of the event-by-event distribution of the mean $p_T$
is in good agreement with predictions based on noncritical thermodynamic 
fluctuations.\footnote{This width can be measured even
if one observes only two pions per event \cite{bialaskoch};
large acceptance data as from NA49 is required in order
to learn that the distribution is Gaussian, that 
thermodynamic predictions may be valid, and that
the width is therefore the only interesting quantity to measure.}
The data on multiplicity fluctuations show evidence for
a nonthermodynamic contribution, which is to 
be expected since the extensive quantity $N$ is sensitive
to the initial size of the system and thus to nonthermodynamic
effects like variation in impact parameter. The
contribution of such effects to the fluctuations
have now been estimated \cite{BH,DS}; the combined thermodynamic
and nonthermodynamic fluctuations are in satisfactory agreement
with the data \cite{DS}.
We use our analysis
to argue \cite{SRS2} that NA49 data are consistent with the hypothesis
that almost all the observed event-by-event fluctuation in 
mean $p_T$, an intensive quantity,
is thermodynamic in origin.
This bodes well for the detectability of systematic
changes in thermodynamic fluctuations near $E$.

As one example, consider
the ratio of
the width of the true event-by-event distribution of 
the mean $p_T$ to the 
width of the distribution in a sample of mixed events \cite{SRS2}. 
We called this
ratio $\sqrt{F}$. NA49
has measured $\sqrt{F}=1.002\pm 0.002$ \cite{NA49,SRS2},
which is consistent with
expectations for noncritical thermodynamic 
fluctuations.\footnote{In an infinite system made of
classical particles which is in thermal equilibrium, $\sqrt{F}=1$.
Bose effects increase $\sqrt{F}$ by $1-2\%$ \cite{Mrow,SRS2}; 
an anticorrelation introduced
by energy conservation in a finite system --- when one mode fluctuates
up it is more likely for others to fluctuate down --- decreases
$\sqrt{F}$ by $1-2\%$ \cite{SRS2}; 
two-track resolution also decreases $\sqrt{F}$
by $1-2\%$ \cite{NA49}. The contributions due to correlations
introduced by resonance decays and due to fluctuations in
the flow velocity are each significantly smaller than $1\%$ \cite{SRS2}.}
We argue \cite{SRS2} that critical fluctuations can increase $\sqrt{F}$
by $10-20\%$, fifty times the statistical error in the present
measurement.   We have focussed on this observable because
data on it has been analyzed and presented by NA49. 
There are, however, other observables which are much more
sensitive to critical effects.  For example, a $\sqrt{F_{\rm soft}}$,
defined using only the softest $10\%$ of the pions in each event, may
easily be affected at the factor of two level.  

We have estimated
the magnitude of the effect of critical fluctuations on
many other observables \cite{SRS2}, including the event-by-event
fluctuations in $N$, and the correlation between fluctuations
in $p_T$ and $N$. We also note one observable,
the multiplicity of soft pions, which may
be used to detect the critical fluctuations 
without an event-by-event analysis.
The post-freezeout decay of sigmas, which are copious
and light at freezeout near $E$ and which
decay subsequently when their mass increases above
twice the pion mass, should result in a population of pions 
with $p_T\sim m_\pi/2$ which appears only for freezeout
near the critical point \cite{SRS2}.  The variety of observables
which should all show nonmonotonic behavior near the critical
point is sufficiently great that if it were to turn out that 
$\mu_E<200$ MeV, making $E$ inaccessible to the SPS, all four
RHIC experiments could play a role in the study of the critical
point.

NA49 data demonstrates very clearly that
SPS collisions at $\sqrt{s}=17$ GeV {\it do not} freeze out
near the critical point.  
$E$ has not yet been discovered.
The nonmonotonic appearance and then disappearance (as 
$\sqrt{s}$ is varied)
of any one of the signatures of the critical fluctuations
we have described would be strong evidence for 
critical fluctuations.  
If nonmonotonic variation is seen
in several of these observables, with the maxima in all
signatures occurring at the same value of $\sqrt{s}$, this
would turn strong evidence into an unambiguous discovery
of the critical point. The quality of the present NA49 data, 
and the confidence with
which we can use it to learn that 
collisions at $\sqrt{s}=17$ GeV do not freeze out
near the critical point
make us confident that the program
of experimentation which we have described can realistically
be expected to teach us much about the phase diagram of QCD,
and could result in a discovery of perhaps the most 
fundamental feature of the landscape.
Once the critical point $E$ is
discovered, it will be prominent on the map of the 
phase diagram which will appear in any future textbook on QCD.

\section{Color Superconductivity and Color-Flavor Locking}

As in my talk in Torino, I devote the final Section of this
review
to recent developments in our understanding
of the low temperature, high density
regions of the QCD phase diagram.  
In this regime, the relevant degrees
of freedom are those which involve quarks with momenta
near the Fermi surface.  At high density, when the 
Fermi momentum is large, the QCD gauge coupling $g(\mu)$ is small.
However, because of the infinite degeneracy among  
pairs of quarks with equal and opposite momenta
at the Fermi surface, 
even an arbitrarily weak attraction between quarks which 
allows a pair of quarks with momenta $\pm \vec p$ to scatter into
a pair with momenta $\pm \vec q$ renders
the Fermi surface unstable to the formation of a condensate
of quark Cooper pairs.  Pairs of quarks cannot be color
singlets, and in QCD with two flavors of massless quarks
the Cooper pairs form in the color ${\bf \bar 3}$ 
channel \cite{Barrois,BailinLove,ARW1,RappETC}.
The resulting condensate gives gaps to quarks with two
of three colors and breaks $SU(3)_{\rm color}$ to an 
$SU(2)_{\rm color}$ subgroup, giving mass to five of
the gluons by the Anderson-Higgs mechanism.
In QCD with two flavors, the Cooper pairs are $ud-du$
flavor singlets and the global flavor symmetry 
$SU(2)_L\times SU(2)_R$ is intact. There
is also an unbroken global symmetry which plays the
role of $U(1)_B$. Thus, no global symmetries are broken
in this 2SC phase.  There must therefore be a phase
transition between the 2SC and hadronic
phases on the horizontal axis in Figure 1, at which
chiral symmetry is restored.  This phase transition
is first order \cite{ARW1,bergesraj,PisarskiRischke1OPT,CarterDiakonov}
as is to be
expected as it 
is characterized by a competition between chiral condensation and
diquark condensation \cite{bergesraj,CarterDiakonov}.
There need be no transition between the 2SC and quark-gluon
plasma phases in Figure 1 because neither phase breaks any global
symmetries.
However, this transition, which
is second order in mean field theory, is likely first
order in QCD \cite{bergesraj}.

In QCD with three flavors of massless quarks, the Cooper
pairs {\it cannot} be flavor singlets, and both color and flavor
symmetries are necessarily broken. The symmetries of
the phase which results have been analyzed 
in \cite{CFL,SW1}.  The attractive channel favored
by one-gluon exchange exhibits ``color-flavor locking.''
A condensate involving left-handed quarks alone 
locks $SU(3)_L$ flavor rotations to $SU(3)_{\rm color}$,
in the sense that the condensate is not symmetric under either alone, but is
symmetric under the 
simultaneous $SU(3)_{L+{\rm color}}$ rotations.\footnote{It turns
out \cite{CFL} that condensation in the color ${\bf \bar 3}$ channel
induces a condensate in the color ${\bf 6}$ channel 
because this breaks no further symmetries \cite{ABR}.
The resulting condensates can be written in terms 
of $\kappa_1$ and $\kappa_2$ where
$\langle q^\alpha_{La} q^\beta_{Lb} \rangle \sim \kappa_1 \delta^\alpha_a
\delta^\beta_b + \kappa_2 \delta^\alpha_b \delta^\beta_a$. Here,
$\alpha$ and $\beta$ are color indices running from $1$ to $3$, $a$ and
$b$ are flavor indices running from $1$ to $3$, and the Kronecker $\delta$'s
lock color and flavor rotations.}
A condensate involving right-handed quarks alone 
locks $SU(3)_R$ flavor rotations to $SU(3)_{\rm color}$.
Because color is vectorial, the result is to lock $SU(3)_L$
to $SU(3)_R$, breaking chiral symmetry.\footnote{Once 
chiral symmetry is broken by color-flavor locking, 
there is no symmetry argument precluding the existence
of an ordinary chiral condensate. Indeed,
instanton effects do induce a nonzero $\langle \bar q q \rangle$ \cite{CFL},
 but this is a small effect \cite{RappETC2}.}
Thus, in quark matter with three massless quarks,
the $SU(3)_{\rm color}\times SU(3)_L \times SU(3)_R \times U(1)_B$
symmetry is broken down to the global diagonal $SU(3)_{{\rm color}+L+R}$
group.  All nine quarks have a gap. All eight gluons get a mass.
There are nine massless Nambu-Goldstone bosons.
There is an unbroken gauged $U(1)$ symmetry which plays 
the role of electromagnetism.
(Under this symmetry, all the quarks, all the massive vector bosons, and
all the Nambu-Goldstone bosons have integer charges.)
The CFL phase therefore has the same symmetries
as baryonic matter with a condensate of
Cooper pairs of baryons \cite{SW1}.  Furthermore,
many non-universal features of these two phases 
correspond \cite{SW1}.
This raises the possibility that quark matter and baryonic
matter may be continuously connected \cite{SW1}, 
as shown in Figure 4.  

The physics of 
the CFL phase has been the focus of much recent 
work \cite{CFL,SW1,ABR,SW2,RappETC2,Zahed,effectiveCFL,gapless}.
Nature chooses two light quarks and one middle-weight
strange quark, rather than three
degenerate quarks as in Figure 4. 
A nonzero $m_s$ weakens those condensates which
involve pairing between light and strange quarks.
The CFL phase requires
nonzero $\langle us \rangle$ and $\langle ds \rangle$
condensates; because these condensates
pair quarks with differing Fermi momenta 
they can only exist if they are 
larger than of order $m_s^2/2\mu$, the
difference between the $u$ and $s$ Fermi momenta in
the absence of pairing.
If one imagines increasing $m_s$ at fixed $\mu$, one finds a first order
unlocking transition \cite{ABR,SW2}: for larger
$m_s$ only $u$ and $d$
quarks pair and the 2SC phase is obtained.  
Conversely, as $m_s$
is reduced in going from Figure 2 to 3 to 4, the 
region occupied by the CFL phase expands to encompass smaller
and smaller $\mu$ \cite{ABR,SW2}.  For  
any $m_s\neq \infty$,
the CFL phase is the ground state at arbitrarily
high density \cite{ABR}.  For larger values of $m_s$,
there is a 2SC interlude on the horizontal axis,
in which chiral symmetry is restored, before
the CFL phase breaks it again at high densities.
For smaller values of $m_s$, the possibility of
quark-hadron continuity as shown in Figure 4 arises.
It should be noted that when the strange and light quarks 
are not degenerate,
the CFL phase may be continuous
with a baryonic phase in which the densities of
all the nucleons and hyperons are comparable; there
are, however, phase transitions between this 
hypernuclear phase and ordinary nuclear matter \cite{ABR}.

The Nambu-Goldstone bosons in the CFL phase
are Fermi surface excitations in which the orientation
of the left-handed and right-handed diquark condensates oscillate
out of phase in flavor space.
The effective field theory describing
these oscillations has recently been constructed \cite{effectiveCFL}.
This effective theory will be useful for many purposes,
including for answering the question posed in Ref. \cite{Zahed}:
what are the properties of the solitons, if any, in the CFL phase?
The dispersion 
relations describing the quasiparticle excitations in
the CFL phase have also received attention \cite{ABR,gapless}.
One interesting possibility is that at the lowest densities
at which the CFL phase exists, just above the first order
unlocking transition between CFL and 2SC, the CFL phase
may be a superconductor in the sense that all eight
gluons are given a mass by the Meissner effect, but there
may nevertheless be gapless quasiparticle excitations \cite{gapless}.

Much effort has gone into estimating
the magnitude of the condensates in the 2SC and CFL 
phases \cite{BailinLove,ARW1,RappETC,CFL,ABR,SW2,bergesraj,CarterDiakonov,RappETC2,Hsu1,SW0,AKS,PisarskiRischke,Son,Hong,HMSW,SW3,PR,rockefeller,Hsu2}.
It would be ideal if this task were within the scope of
lattice gauge theory as is, for example, the calculation
of the critical temperature on the vertical axis of the phase diagram.
Unfortunately, lattice methods relying
on importance sampling have to this
point been rendered exponentially 
impractical at nonzero baryon density by the 
complex action at nonzero $\mu$.\footnote{Note that 
quark pairing can be studied on the lattice in some models
with four-fermion interactions and in two-color QCD \cite{HandsMorrison}. 
The $N_c=2$ case has also
been studied analytically in Refs. \cite{RappETC,analytic2color}; 
pairing in this
theory is simpler to analyze because 
quark Cooper pairs are color singlets. 
The $N_c\rightarrow \infty$
limit of QCD is often one in which hard problems become
tractable. However, the ground state of $N_c=\infty$ QCD
is {\it not} a color superconductor \cite{DGR}. This is
of no concern at $N_c=3$, however: Shuster and Son \cite{eugene}
have shown that color superconductivity persists up
to $N_c$'s of order tens of thousands before being
supplanted by the phase described in Ref. \cite{DGR}.}
In the absence of lattice methods, the
magnitude of the gaps in quark matter at large but accessible
density has been estimated using two broad strategies.
The first class of estimates are done within the
context of models whose
parameters are chosen to give reasonable vacuum
physics.  Examples include analyses in which the
interaction between quarks is replaced simply by four-fermion
interactions with the quantum numbers of
the instanton interaction \cite{ARW1,RappETC,bergesraj}
or of one-gluon exchange \cite{CFL,ABR} and more sophisticated
analyses done using
the instanton liquid model \cite{CarterDiakonov,RappETC2}.
Renormalization group methods have also been used to
explore the space of all possible 
effective four-fermion interactions \cite{Hsu1,SW0}.
These methods yield results which are in qualitative 
agreement: the favored condensates are as described
above; the gaps range between several tens of MeV up to as large as about 
$100$ MeV;
the associated critical temperatures (above which the 
diquark condensates vanish)
can be as large as about $T_c\sim 50$ MeV.
This agreement between different models reflects the fact
that what matters most is simply the strength of the attraction
between quarks in the color ${\bf \bar 3}$ channel, and by
fixing the parameters of the model interaction to fit, say,
the magnitude of the  vacuum chiral condensate, one ends up
with attractions of similar strengths in different models.

The second strategy for estimating gaps and critical
temperatures is to use
$\mu=\infty$ physics as a guide.
At asymptotically large $\mu$, models with short-range interactions
are bound to fail because the dominant interaction is due
to the long-range magnetic interaction coming from single-gluon
exchange \cite{PisarskiRischke,Son}.  The collinear infrared
divergence in small angle scattering via one-gluon exchange
(which is regulated by dynamical screening \cite{Son})
results in a gap which is parametrically larger at $\mu\rightarrow\infty$
than it would be for any point-like four-fermion interaction.
At $\mu\rightarrow\infty$, where $g(\mu)\rightarrow 0$,
the gap takes the 
form $\Delta \sim 256 \pi^4 \mu \,g(\mu)^{-5} \exp[-3\pi^2/\sqrt{2}g(\mu)]$;
the $g(\mu)$ dependence was discovered in Ref. \cite{Son} and
has now been confirmed using a variety of 
methods \cite{SW3,PR,rockefeller,Hsu2}.  A consequence
of this result is that the magnitude of the condensates 
increases slowly as $\mu\rightarrow\infty$; this means
that the CFL phase is favored over the 2SC phase 
for $\mu\rightarrow\infty$ for any $m_s\neq \infty$ \cite{ABR}.
The $g$-independent prefactor
has been estimated in Refs. \cite{SW3,PR,rockefeller}.
As conjectured in \cite{ABR}, this prefactor is such that
if this asymptotic expression is applied to accessible densities,
say $\mu\sim 500 - 1000$ MeV, it predicts gaps as large as 
about $100$ MeV and
critical temperatures as large as about $50$ MeV \cite{SW3}. 
The weak-coupling calculation of the gap in the CFL phase is the
first step toward the weak-coupling calculation of other
properties of this phase, in which chiral symmetry is broken
and the spectrum of excitations is as in a confined phase,
like for example the coefficients in the effective field
theory of Ref. \cite{effectiveCFL} which describes the physics
of the Nambu-Goldstone bosons.

It is satisfying that two very different approaches,
one using zero density phenomenology to normalize models, the
other using weak-coupling methods valid at asymptotically
high density, yield predictions for the
gaps and critical temperatures at accessible
densities
which are in good agreement.
$T_c\sim 50$ MeV is much larger relative to the
Fermi momentum (say $\mu\sim 500$ MeV) than in 
low temperature superconductivity in metals.
This reflects the fact that color superconductivity
uses attraction due to the primary,
strong, interaction in the theory, rather
than having to rely on much weaker secondary interactions,
as in phonon mediated superconductivity in metals.
Quark matter is a high-$T_c$ superconductor by any reasonable
definition. It is unfortunate
that its $T_c$ is nevertheless low enough that
it is unlikely the phenomenon can be realized in heavy ion
collisions.  

Neutron stars have $T\sim 0$, relative to the scales of Figures 1-4.
If they have quark matter cores, those cores are color
superconductors, in either the 2SC or CFL phase. 
The higher temperature regions of the phase diagram are
being mapped
in heavy ion collisions; we need to learn
how to use neutron star phenomenology to  map the
high density regime.  The qualitative questions
to which analysis of neutron star phenomenology may yet provide
answers include: Does baryonic matter change
continuously into quark matter in the CFL
phase as a function of increasing density (increasing depth)
within a neutron star?
This would require that the high density region of the phase
diagram be similar to Figure 4.
Or, as in Figure 3, do neutron stars have a 2SC core, perhaps
with a CFL inner core?  Or, of course, do neutron stars not
have quark matter cores at all?  
Further work is required.
The rate of cooling of neutron stars by neutrino
emission from their cores may yield useful
information: both the emission rates and
the heat capacity are affected by diquark condensates \cite{schaab}
and the cooling rates resulting in the  
2SC and CFL phases differ \cite{blaschke}.  The $r$-mode instability
is sensitive to the presence of quark matter within neutron 
stars \cite{madsen}; it remains to be seen whether it differentiates
between the 2SC and CFL phases.  Pulsars are characterized by
strong magnetic fields; 
further work is required to determine how
the field configuration would be
affected by the presence of a 2SC or CFL core \cite{flux},
and whether the time evolution of pulsar magnetic fields
would be affected.

The answer to the question of whether the QCD phase diagram 
does or does
not feature a 2SC interlude
on the horizontal axis, separating the CFL and baryonic phases
in both of which chiral symmetry is broken,
depends on whether the strange
quark is effectively heavy or effectively light.  This is the
central outstanding question about how to draw
the high density region of the map.
The central question at higher temperatures, 
namely where does nature locate
the critical
point $E$, also depends on the strange quark mass. 
Both questions are hard to answer theoretically with
any confidence.  The higher temperature region of the map
is in much better shape, however, because the program of experimentation
described in the previous Section allows 
heavy ion collision experiments to search for 
the critical point $E$. 
Theorists have described how to use phenomena characteristic of freezeout
in its vicinity to
discover $E$;  this enables experimentalists
to answer the question of its location convincingly.
The discovery of $E$ would allow us to draw the higher temperature
regions of the map of the QCD phase diagram in ink.
Theorists have much work to do, however, before a  
program of astrophysical observations which
would allow the inking in of the boundaries of the 2SC and CFL
phases can be proposed.

\end{document}